\documentclass[11pt]{article}
\usepackage{epsfig}
\usepackage{rotating}

\newcommand{\BABARPubYear}    {07}

\newcommand{\BABARConfNumber} {009}
\newcommand{\SLACPubNumber} {12735}

\input babarsym

\long\def\inst#1{\par\nobreak\kern 4pt\nobreak
    {\it #1}\par\vskip 10pt plus 3pt minus 3pt}

\begin{document}
{\pagestyle{empty}


\begin{flushright}
\babar-CONF-\BABARPubYear/\BABARConfNumber \\
SLAC-PUB-\SLACPubNumber \\
August 2007 \\
\end{flushright}

\par\vskip 4cm

\begin{center}
\Large \bf Measurement of the Branching Fractions of Exclusive {\boldmath $\overline{B} \to D/D^{*}/D^{(*)}\pi \ell^- \bar{\nu}_{\ell}$} Decays in Events Tagged by a Fully Reconstructed {\boldmath $B$} Meson
\end{center}
\bigskip

\begin{center}
\large The \babar\ Collaboration\\
\mbox{ }\\
\today
\end{center}
\bigskip \bigskip

\begin{center}
\large \bf Abstract
\end{center}
We present a measurement of the branching fractions for $\overline{B} \to D/D^{*}/D^{(*)}\pi \ell^- \bar{\nu}_{\ell}$ decays based on 
341.1 fb$^{-1}$ of data collected at the $\Upsilon(4S)$ resonance with the \babar\
detector at the \pep2\ $e^+e^-$ storage rings. 
Events are tagged by fully reconstructing one of the $B$ mesons in a hadronic decay mode. We obtain ${\cal B} (B^- \to D^0 \ell^- \bar{\nu}_{\ell}) = (2.33 \pm 0.09~(\mbox{stat.}) \pm 0.09~(\mbox{syst.}))\%$, ${\cal B} (B^- \to D^{*0} \ell^- \bar{\nu}_{\ell}) = (5.83 \pm 0.15~(\mbox{stat.}) \pm 0.30~(\mbox{syst.}))\%$, ${\cal B} (\overline{B^0} \to D^+ \ell^- \bar{\nu}_{\ell}) = (2.21 \pm 0.11~(\mbox{stat.}) \pm 0.12~(\mbox{syst.}))\%$, ${\cal B} (\overline{B^0} \to D^{*+} \ell^- \bar{\nu}_{\ell}) = (5.49 \pm 0.16~(\mbox{stat.}) \pm 0.25~(\mbox{syst.}))\%$, ${\cal B} (B^- \to D^+ \pi^- \ell^- \bar{\nu}_{\ell}) = (0.42 \pm 0.06~(\mbox{stat.}) \pm 0.03~(\mbox{syst.}))\%$, ${\cal B} (B^- \to D^{*+}\pi^- \ell^- \bar{\nu}_{\ell}) = (0.59 \pm 0.05~(\mbox{stat.}) \pm 0.04~(\mbox{syst.}))\%$, ${\cal B} (\overline{B^0} \to D^0 \pi^+ \ell^- \bar{\nu}_{\ell}) = (0.43 \pm 0.08~(\mbox{stat.}) \pm 0.03~(\mbox{syst.}))\%$ and ${\cal B} (\overline{B^0} \to D^{*0} \pi^+ \ell^- \bar{\nu}_{\ell}) = (0.48 \pm 0.08~(\mbox{stat.}) \pm 0.04~(\mbox{syst.}))\%$. 
\vfill
\begin{center}
Contributed to the
XXIII$^{\rm rd}$ International Symposium on Lepton and Photon Interactions at High~Energies, 8/13 -- 8/18/2007, Daegu, Korea
\end{center}

\vspace{1.0cm}
\begin{center}
{\em Stanford Linear Accelerator Center, Stanford University,
Stanford, CA 94309} \\ \vspace{0.1cm}\hrule\vspace{0.1cm}
Work supported in part by Department of Energy contract DE-AC03-76SF00515.
\end{center}

\newpage
} 

%
%
\begin{center}
\small

The \babar\ Collaboration,
\bigskip

%
{B.~Aubert,}
{M.~Bona,}
{D.~Boutigny,}
{Y.~Karyotakis,}
{J.~P.~Lees,}
{V.~Poireau,}
{X.~Prudent,}
{V.~Tisserand,}
{A.~Zghiche}
\inst{Laboratoire de Physique des Particules, IN2P3/CNRS et Universit\'e de Savoie, F-74941 Annecy-Le-Vieux, France }
{J.~Garra~Tico,}
{E.~Grauges}
\inst{Universitat de Barcelona, Facultat de Fisica, Departament ECM, E-08028 Barcelona, Spain }
% ...

\end{center}\newpage

\section{Introduction}
\label{sec:Introduction}

The determination of the individual exclusive branching fractions of
$\overline{B} \to X_c \ell^- \bar{\nu}_{\ell}$ decays\footnote{Here $X_c$ refers to any charm hadronic state, $X_u$ to any charmless hadronic state,  and $\ell = e, \mu$.} is important for the study of the
dynamics of semileptonic decays of the $B$ meson. 
Improvement in the knowledge  of these branching fractions is also desired to reduce the systematic uncertainty in the measurements of the Cabibbo-Kobayashi-Maskawa~\cite{CKM} elements $|V_{cb}|$ and $|V_{ub}|$. This is exemplified by the fact that one of the leading sources of
systematic uncertainty in the extraction of $|V_{cb}|$ from the exclusive
decay $\overline{B} \to D^* \ell^- \bar{\nu}_{\ell}$ is our limited knowledge of the
background due to $\bar{B} \to D^* \pi \ell^- \bar{\nu}_{\ell}$~\cite{pdg}. Improvements in our knowledge of $\overline{B} \to X_c \ell^- \bar{\nu}_{\ell}$ decays will also benefit the accuracy of the extraction of $|V_{ub}|$, as analyses are
extending into kinematical regions where these decays represent a sizable background.

Measurements of the largest $B$ meson branching fraction, ${\cal B} (\overline{B^0} \to D^{*+} \ell^- \bar{\nu}_{\ell})$, by the CLEO, LEP, \babar\,, and BELLE experiments, based  on different reconstruction techniques, show sizable differences, resulting in 
an average value~\cite{hfag} with a probability of only 
5\%, thus indicating uncertainties that may be larger than stated and/or sizable 
correlations among the different measurements. The world average for the charged mode, ${\cal B} (B^- \to D^{*0} \ell^- \nu_{\ell})$, has a larger uncertainty (about 8\%) and is dominated by the CLEO result~\cite{cleoDs0}. Recently \babar\ has measured the form-factor parameters and the branching fraction of the $B^- \to D^{*0} \ell^- \nu_{\ell}$ decay mode, with an accuracy comparable to the world average~\cite{schubert}. 
Few measurements are available for the exclusive decays $\overline{B} \to D \ell^- \bar{\nu}_{\ell}$ and the uncertainties are in general larger due to larger background levels and less stringent kinematic constraints (i.e. no narrow $D^{*}$ mass peak)~\cite{pdg}. 

The first measurements of the fractions of $D^{**}$ states in semileptonic
$B$ decays were reported by the ALEPH~\cite{Buskulic:1996uk},
DELPHI~\cite{Abreu:2000,Bloch:2000}, and OPAL~\cite{Abbiendi:2002ge}
collaborations at LEP, and by CLEO~\cite{Anastassov:1997im} at CESR.  We denote by $D^{**}$ any hadronic final
state containing a charm meson with total mass above that of the $D^*$
state, thereby including both  $D_J$ excited mesons (narrow and broad states) and $D^{(*)}$+$n \pi$ non-resonant
 states, with $n \geq 1$. More recently, new
results have been obtained by the D0~\cite{Abazov:2005ga}
experiment at the Tevatron.

While $\overline{B} \rightarrow D \ell^- \bar{\nu}_{\ell}$ and  $\overline{B} \rightarrow D^* \ell^- \bar{\nu}_{\ell}$ decays account for
about 70\% of the $\overline{B} \to X \ell^- \bar{\nu}_{\ell}$ branching fraction, the contributions of other states, including resonant and non-resonant $D^{(*)}n\pi$ decays, 
are not yet well measured and are a possible explanation of the observed difference between the inclusive rate and the  sum of the measured exclusive rates.
BELLE has recently reported the determination of the
$D^{(*)} \pi \ell \bar{\nu}_{\ell}$ decay branching fraction~\cite{Abe:2005up}, and \babar\ \cite{babarmio} has reported a measurement of the relative branching fractions $\Gamma(\overline{B} \rightarrow D/D^{*}/D^{**} \ell^- \bar{\nu}_{\ell})/\Gamma (\overline{B} \rightarrow D X \ell^- \bar{\nu}_{\ell})$. 
In this paper, we present measurements of the branching fractions for $\overline{B} \to D/D^{*}/D^{(*)}\pi \ell^- \bar{\nu}_{\ell}$  decays\footnote{Charge-conjugate modes are implied throughout this paper, unless explicitly stated otherwise.}, separately for charged and neutral $B$ mesons. These exclusive branching fractions are normalized to a sample of inclusive $\overline{B} \to X \ell^- \bar{\nu}_{\ell}$ events.

\section{The \babar\ Detector and Dataset}
\label{sec:babar}
This analysis is based on data collected with the \babar\ detector at the \pep2\ storage rings. The total integrated luminosity of the dataset is 341.1 fb$^{-1}$ collected on the $\Upsilon(4S)$ resonance. The corresponding number of produced \BB\ pairs is 378 million. An additional 36 fb$^{-1}$  off-peak data sample taken at a center-of-mass energy 40 MeV below the $\Upsilon(4S)$ resonance is used to study background from $e^+e^- \to f\bar{f}~(f=u,d,s,c,\tau)$ events (continuum production). The \babar\ detector is described in detail elsewhere~\cite{detector}. Charged-particle trajectories are measured by a 5-layer double-sided silicon vertex tracker and a 40-layer drift chamber, both operating in a 1.5-T magnetic field. Charged-particle identification is provided by the average energy loss (d$E$/d$x$) in the tracking devices and by an internally reflecting ring-imaging Cherenkov detector. Photons are detected by a CsI(Tl) electromagnetic calorimeter (EMC). Muons are identified by the instrumented magnetic-flux return. 
A detailed GEANT4-based Monte Carlo (MC) simulation~\cite{Geant} of \BB\ and continuum events has been used to study the detector response, its acceptance, and to test the analysis techniques. The simulation models $\overline{B} \to D^* \ell^- \bar{\nu}_{\ell}$
decays using calculations based on Heavy Quark Effective Theory (HQET) as in~\cite{HQETBaBar}, $\overline{B} \to D \ell^- \bar{\nu}_{\ell}$ and $\overline{B} \to D^{**}(\rightarrow D^{(*)} \pi) \ell^- \bar{\nu}_{\ell}$ decays using the ISGW2 model~\cite{ISGW}, and $\overline{B}
\to D^{(*)} \pi \ell^- \bar{\nu}_{\ell}$ decays using the Goity-Roberts model~\cite{Goity}.

\section{Event Selection}
\label{sec:Analysis}
We select semileptonic $B$-meson decays in events containing
a fully reconstructed $B$ meson ($B_{tag}$), which allows us to constrain the kinematics, to reduce the combinatorial background, and to determine the charge and flavor of the signal $B$.

The analysis exploits the presence of two charmed mesons in the final state:
one used for the exclusive reconstruction of the $B_{tag}$, and the other
for the reconstruction of the semileptonic $B$ decay.

We first reconstruct the semileptonic $B$ decay, selecting a lepton with momentum $p^*_{\ell}$ in the  center-of-mass (CM) frame higher than 0.6 GeV/$c$. Electrons from photon conversion and $\pi^0$ Dalitz decay are removed using a dedicated algorithm, which performs the reconstruction of vertices between tracks of opposite charges whose invariant mass is compatible with a photon conversion or a $\pi^0$ Dalitz decay.  Candidate $D^0$ mesons, with the correct 
charge-flavor correlation with the lepton, are reconstructed
in the $K^-\pi^+$, $K^- \pi^+ \pi^0$, $K^- \pi^+ \pi^+ \pi^-$,
$K^0_S \pi^+ \pi^-$, $K^0_S \pi^+ \pi^- \pi^0$, $K^0_S \pi^0$, $K^+ K^-$,
$\pi^+ \pi^-$, and $K^0_S K^0_S$ channels, and $D^+$ mesons in the
$K^- \pi^+ \pi^+$, $K^- \pi^+ \pi^+ \pi^0$, $K^0_S \pi^+$, $K^0_S \pi^+ \pi^0$,
$K^+ K^- \pi^+$, $K^0_S K^+$, and $K^0_S \pi^+ \pi^+ \pi^-$ channels.
In events with multiple $\overline{B} \to D \ell^- \bar{\nu}_{\ell}$ candidates, the candidate with the largest $D$-$\ell$ vertex fit probability is selected.
Candidate $D^{*+}$ mesons are reconstructed by combining a $D$ candidate and a pion in the $D^{*+} \rightarrow D^0 \pi^+ $ and $D^{*+} \rightarrow D^+ \pi^0$ decays. Candidate $D^{*0}$ mesons are reconstructed by combining a $D$ candidate and a pion or a photon in the $D^{*0} \rightarrow D^0 \pi^0$ and $D^{*0} \rightarrow D^0 \gamma$ decays.
In events with multiple $\overline{B} \to D^{*} \ell^- \bar{\nu}_{\ell}$ candidates, we choose the candidate with the smallest $\chi^2$ based on the deviations from the nominal values of the $D$ invariant mass and the invariant mass difference between the $D^*$ and the $D$, using the measured uncertainty in the mass measurements.

 We reconstruct $B_{tag}$ decays of the type $\overline{B} \rightarrow D Y$, where 
$Y$ represents a collection of hadrons with a total charge of $\pm 1$, composed
of $n_1\pi^{\pm}+n_2 K^{\pm}+n_3 K^0_S+n_4\pi^0$, where $n_1+n_2 \leq  5$, $n_3
\leq 2$, and $n_4 \leq 2$. Using $D^0(D^+)$ and $D^{*0}(D^{*+})$ as seeds for $B^-(\overline{B^0})$ decays, we reconstruct about 1000 types of decay chains.

The kinematic consistency of a $B_{tag}$ candidate with a $B$-meson decay is checked using two variables: the beam-energy
substituted mass $m_{ES}=\sqrt{s/4-\vec{p}_B^2}$, and the energy difference $\Delta E = E_B -\sqrt{s}/2$. Here $\sqrt{s}$ refers to the total CM  energy, and $\vec{p}_B$ and $E_B$ denote the momentum and energy of the $B_{tag}$ candidate in the CM frame. For correctly identified $B_{tag}$ decays, the $m_{ES}$ distribution peaks at the $B$ meson mass, while $\Delta E$ is consistent
with zero.
We select a $B_{tag}$ candidate in the signal region
defined as 5.27~GeV/$c^2$ $< m_{ES} <$ 5.29~GeV/$c^2$, excluding $B_{tag}$ candidates with
 daughter particles in common with the charm meson or
the lepton from the semileptonic $B$ decay. In the case of multiple $B_{tag}$ candidates, we select the one with the smallest
$|\Delta E|$ value. The $B_{tag}$ and the $D^{(*)}\ell$ candidates are required to have the correct charge-flavor correlation. Mixing effects in the $\overline{B^0}$ sample are accounted for as described in~\cite{BBmixing}. 
Cross-feed effects, i.e. $B^-_{tag} (\overline{B^0}_{tag})$ candidates erroneously reconstructed as a neutral~(charged) $B$,  are subtracted using the  MC simulation.

Exclusive $\overline{B} \rightarrow D X \ell^- \bar{\nu}_{\ell}$ decays are identified by relatively loose selection criteria. We require the reconstructed  ground-state charm meson invariant mass $M_{D^0}$ ($M_{D^+}$) to be in the range from 1.850
(1.853)~GeV/$c^2$ to 1.880 (1.883)~ GeV/$c^2$ and the cosine of the angle between the directions of the $D$
candidate and the lepton in the CM frame to be less than zero, to reduce background from non-$B$ semileptonic decays.
We select $\overline{B} \rightarrow D^{*} X \ell^- \bar{\nu}_{\ell}$ decays by requiring the invariant mass difference between the $D^{*}$ and the $D$ to satisfy the selection criteria listed in Table \ref{tab:Dstarcuts}.

\begin{table}[!t]
\caption{Invariant mass ranges for $D^{*0}$ and $D^{*+}$ selection.}
\centering
\begin{tabular}{|cc|}
\hline
\hline
Mode & Selection Criteria \\
\hline
$D^{*0} \rightarrow D^0\pi^0$ &  $0.139 < M(D^{*0}) - M(D^0) < 0.145$ GeV/$c^2$
\\
$D^{*0} \rightarrow D^0\gamma$ &  $0.133 < M(D^{*0}) - M(D^0) < 0.151$ GeV/$c^2$ 
\\
$D^{*+} \rightarrow D^0\pi^+$ &  $0.141 < M(D^{*+}) - M(D^0) < 0.149$ GeV/$c^2$
\\
$D^{*+} \rightarrow D^+\pi^0$ &  $0.138 < M(D^{*+}) - M(D^+) < 0.143$ GeV/$c^2$
\\
\hline
\hline
\end{tabular}
\label{tab:Dstarcuts}
\end{table}

We reconstruct $B^- \rightarrow D^{(*)+}\pi^- \ell^- \bar{\nu}_{\ell}$ and $\overline{B^0} \rightarrow D^{(*)0}\pi^+ \ell^- \bar{\nu}_{\ell}$ decays starting from the corresponding $\overline{B} \rightarrow D^{(*)} X \ell^- \bar{\nu}_{\ell}$ samples and selecting events with only one additional reconstructed charged track that has not been used for the reconstruction of the $B_{tag}$, the signal $D^{(*)}$ or the lepton. 
For the $\overline{B^0} \rightarrow D^0\pi^+ \ell^- \bar{\nu}_{\ell}$ and the $\overline{B^0} \rightarrow D^{*0}\pi^+ \ell^- \bar{\nu}_{\ell}$ decays, we additionally require the invariant mass difference $M(D\pi)-M(D)$ to be greater than 0.18 GeV/$c^2$ to veto $\overline{B^0} \rightarrow D^{*+} \ell^- \bar{\nu}_{\ell}$ events. To reduce the combinatorial background in the $\overline{B^0} \rightarrow D^{*0}\pi^+ \ell^- \bar{\nu}_{\ell}$ mode, we also require the total extra energy in the event, obtained by summing the energy of all the showers in the EMC that have not been assigned to the $B_{tag}$ or the $D^{(*)}\ell^-$ candidates, to be less than 1 GeV.

The semileptonic $B$ decays are identified by the missing mass squared, defined as:

\begin{equation}
 m^2_{miss} = (p(\Upsilon(4S)) -p(B_{tag}) - p(D^{(*)}(\pi)) - p(\ell))^2
\end{equation}

\noindent in terms of the particle
four-momenta in the CM frame of the reconstructed final
states. For correctly reconstructed signal events, the only missing particle is the neutrino, and the $m^2_{miss}$ peaks at zero. Other $B$ semileptonic decays, where one particle is not reconstructed (feed-down) or erroneously added (feed-up) to the charm candidate, spread to higher or  lower values of $m^2_{miss}$. The selection of fully reconstructed
events results in an $m^2_{miss}$ resolution of 0.04~GeV$^2$/$c^4$, an order of magnitude lower compared
to non-tagged analyses~\cite{exampArgus}.

To obtain the $B$ semileptonic signal yields, we perform a one-dimensional extended binned maximum likelihood fit to the $m^2_{miss}$ distributions,  based on a method developed by R.~Barlow and C.~Beeston~\cite{Barlow}. The fitted data samples are assumed to contain four different types of events:

\begin{itemize}
\item signal $\overline{B} \to D/D^{*}/D^{(*)}\pi \ell^- \bar{\nu}_{\ell}$ 
\item feed-down or feed-up $B$ semileptonic decays
\item \BB\ and continuum background
\item fake lepton events. 
\end{itemize}

\noindent For the fit to the $\overline{B} \to D^{(*)}\pi \ell^- \bar{\nu}_{\ell}$ $m^2_{miss}$ distributions, we also include a dedicated component corresponding to other $\overline{B} \to D^{**}(D^{*}\pi) \ell^- \bar{\nu}_{\ell}$ decays different from the exclusive mode being reconstructed. 
We use the Monte Carlo predictions for the different $B$ semileptonic decay $m^2_{miss}$ distributions to obtain the Probability Density Functions (PDFs) to fit the data distributions. The \BB\ and continuum background shape is also estimated by the MC simulation, and we use the off-peak data to provide the continuum background normalization. The shape of the continuum background predicted by the MC simulation is consistent with the one obtained from the off-peak data. 

\section{Measurement of Branching Fractions}

\begin{figure}[!h]
\centering
\epsfig{figure=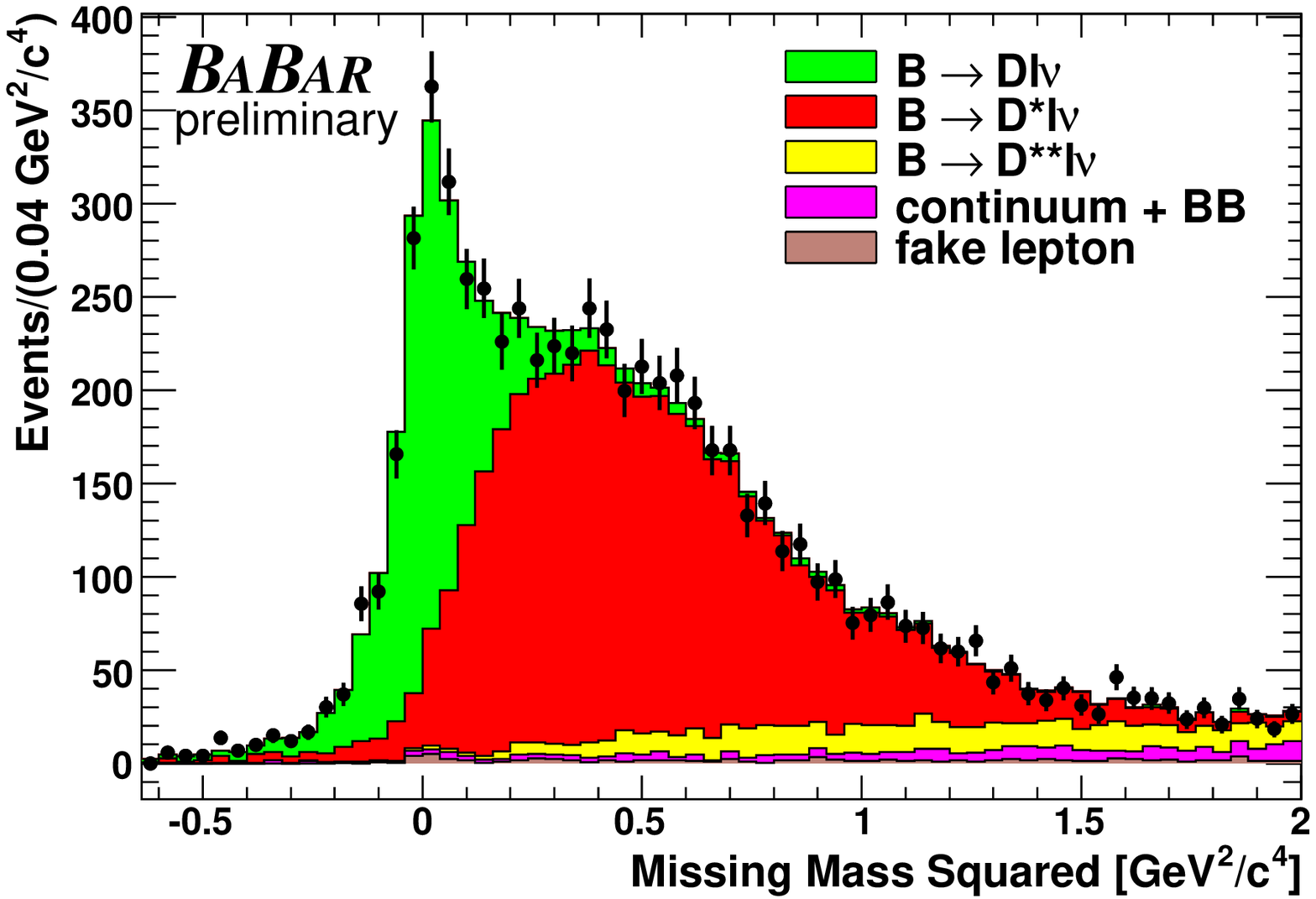,width=8.1cm}
\epsfig{figure=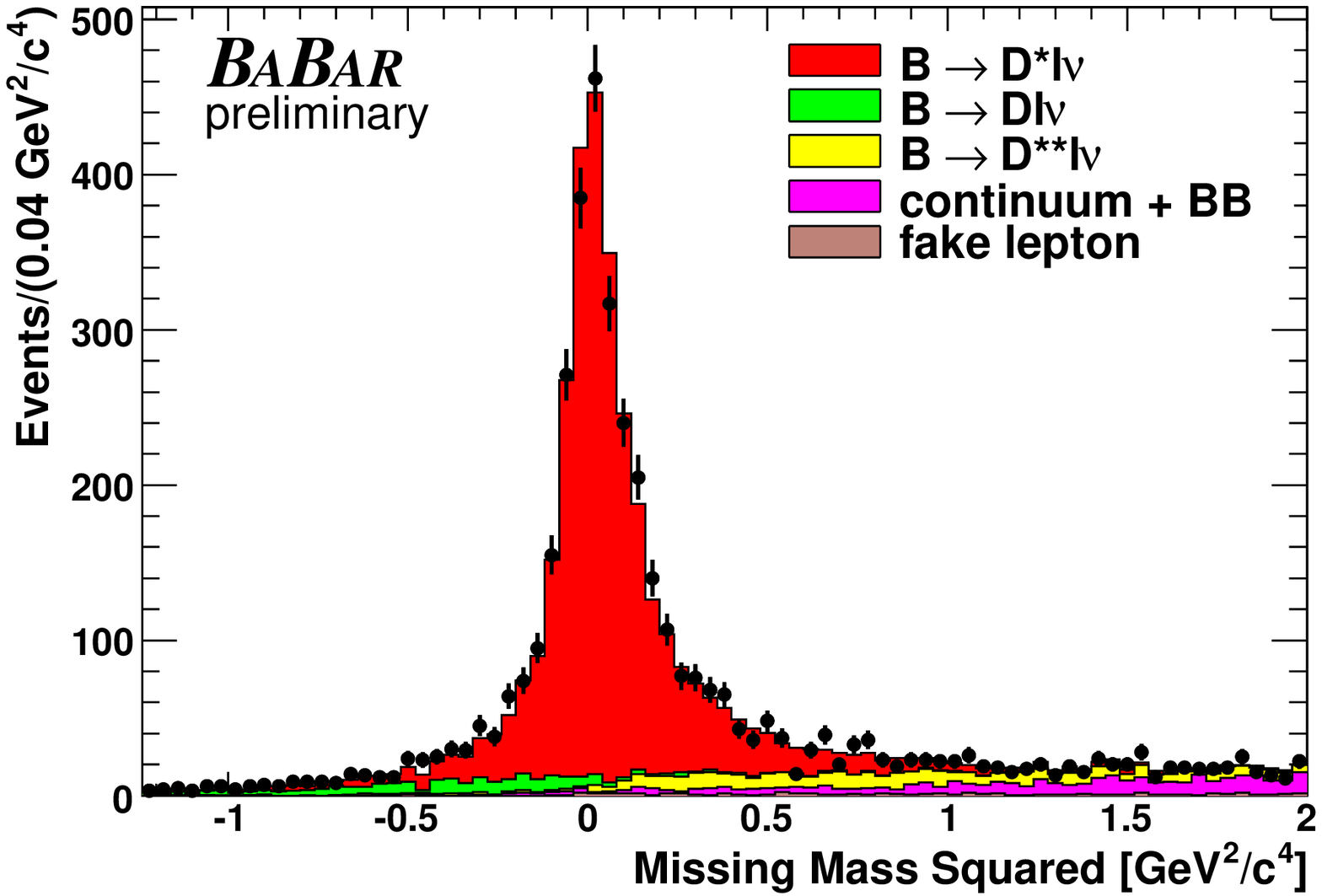,width=8.1cm}
\epsfig{figure=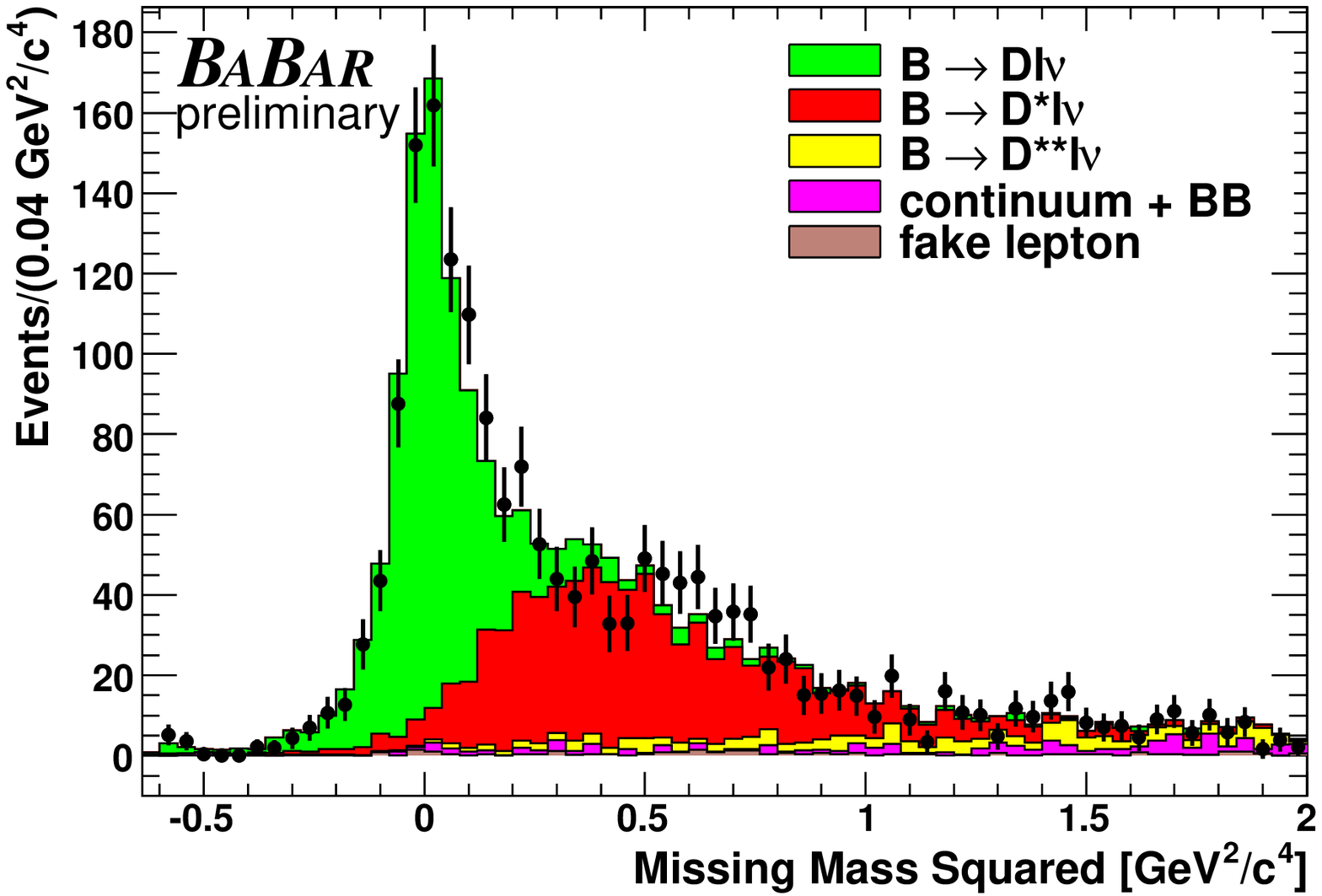,width=8.1cm}
\epsfig{figure=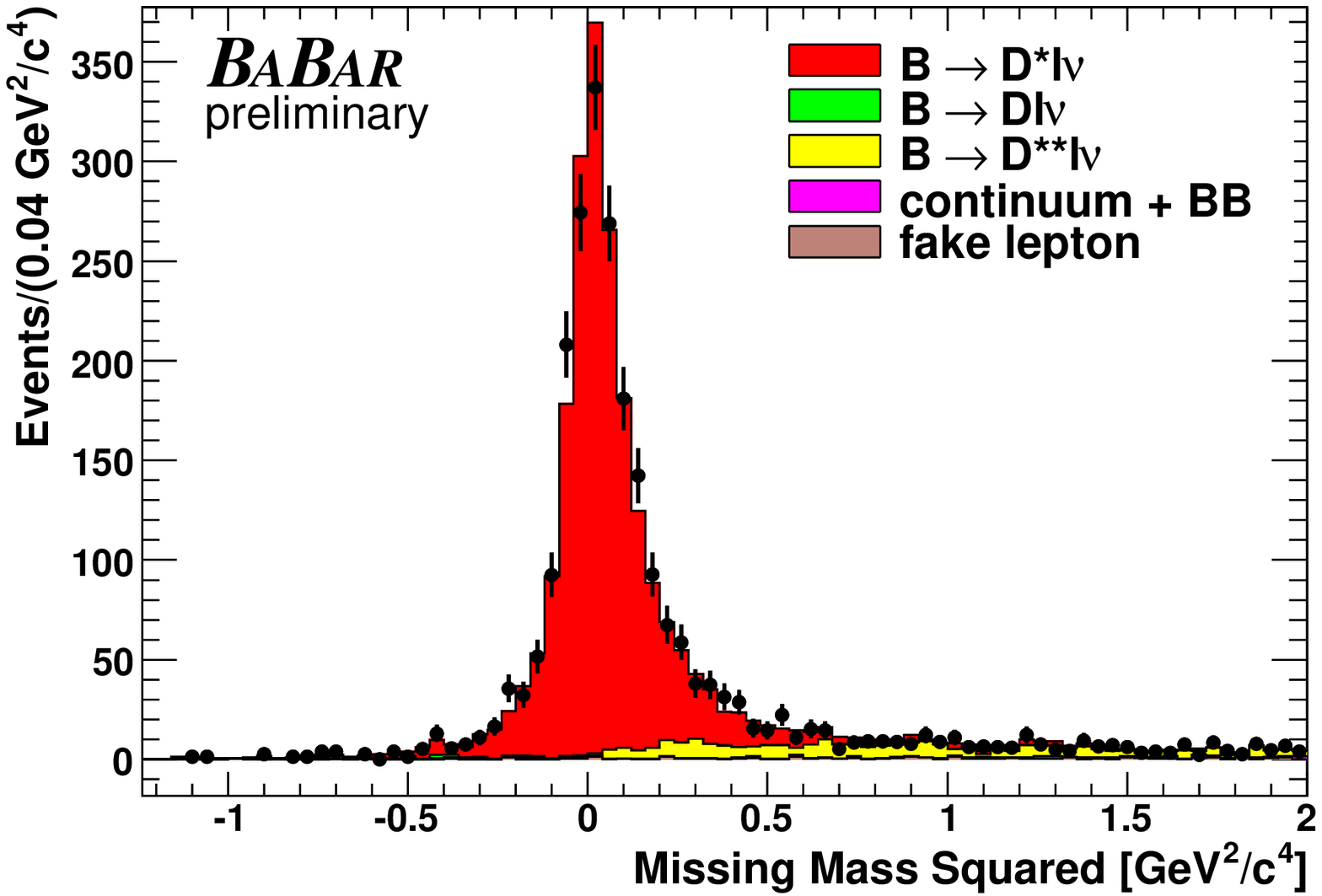,width=8.1cm}
\caption{Fit to the $m^2_{miss}$ distribution for $B^- \to D^0 \ell^- \bar{\nu}_{\ell}$ (top left), $B^- \to D^{*0} \ell^- \bar{\nu}_{\ell}$ (top right),  $\overline{B^0} \to D^+ \ell^- \bar{\nu}_{\ell}$ (bottom left), and $\overline{B^0} \to D^{*+} \ell^- \bar{\nu}_{\ell}$ (bottom right): the data (points with error bars) are compared to the results of the overall fit (sum of the solid histograms). The PDFs for the different fit components are stacked and shown in different colors.}
\label{fig:Fit1}
\end{figure}

\begin{figure}[!h]
\centering
\epsfig{figure=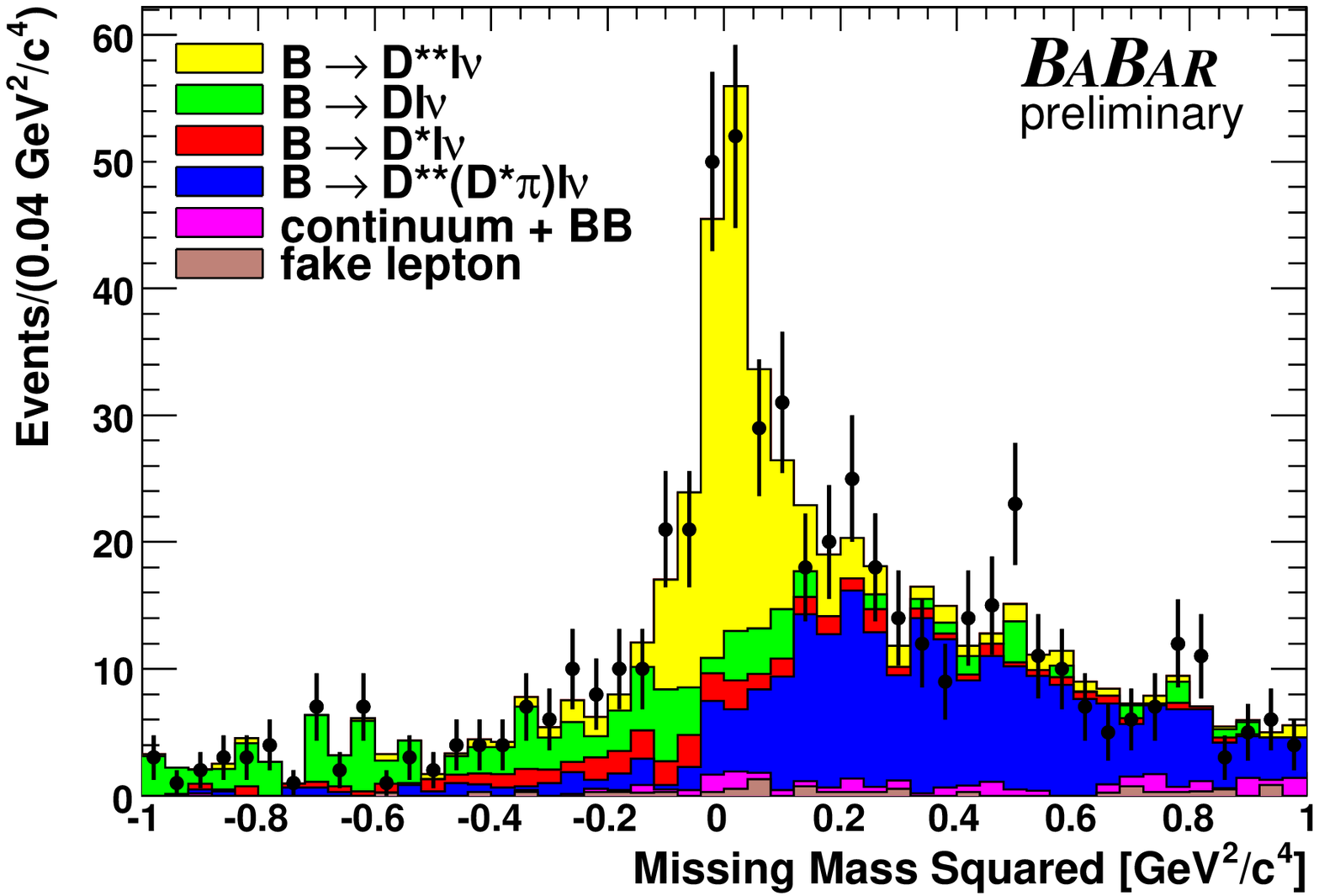,width=8.1cm}
\epsfig{figure=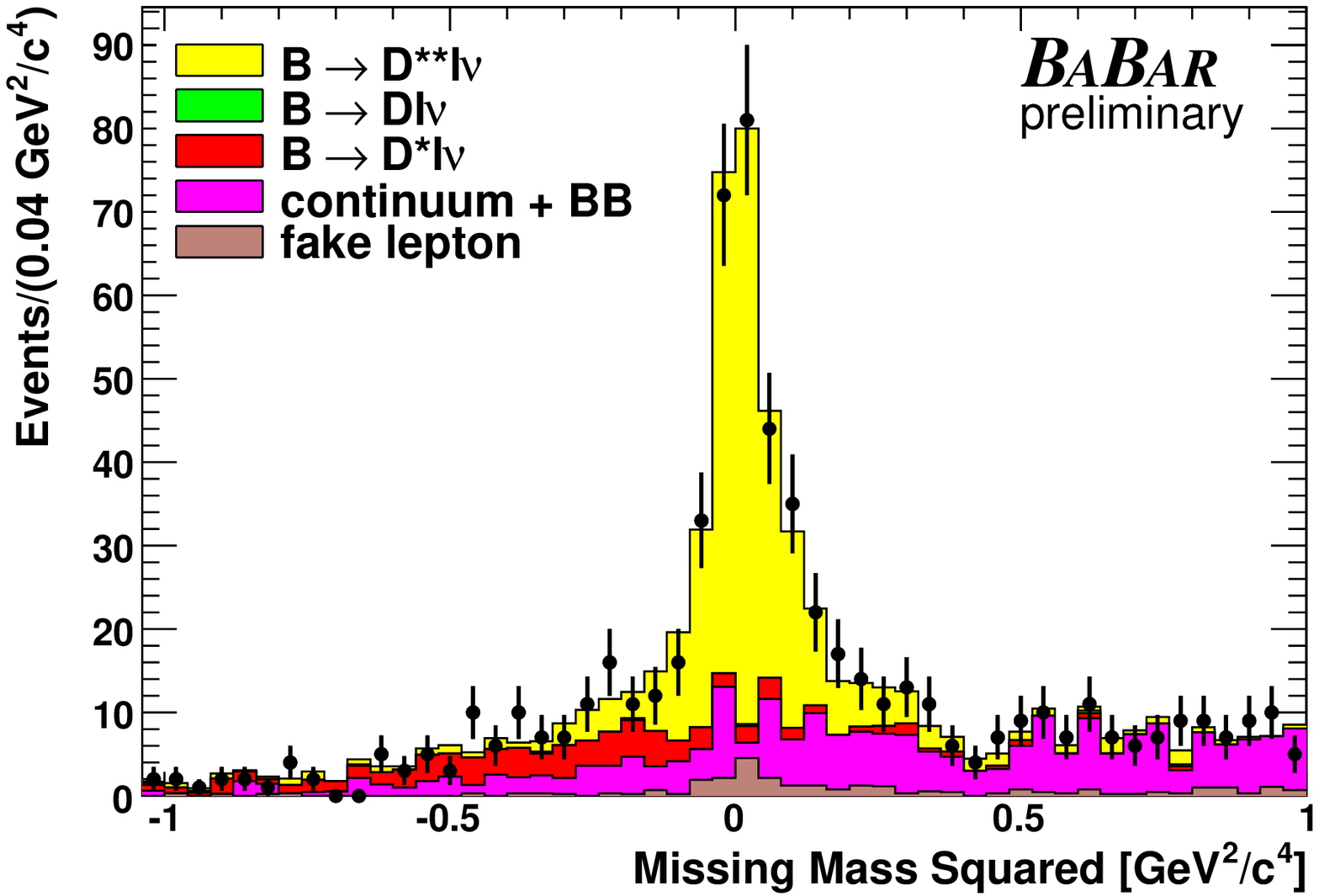,width=8.1cm}
\epsfig{figure=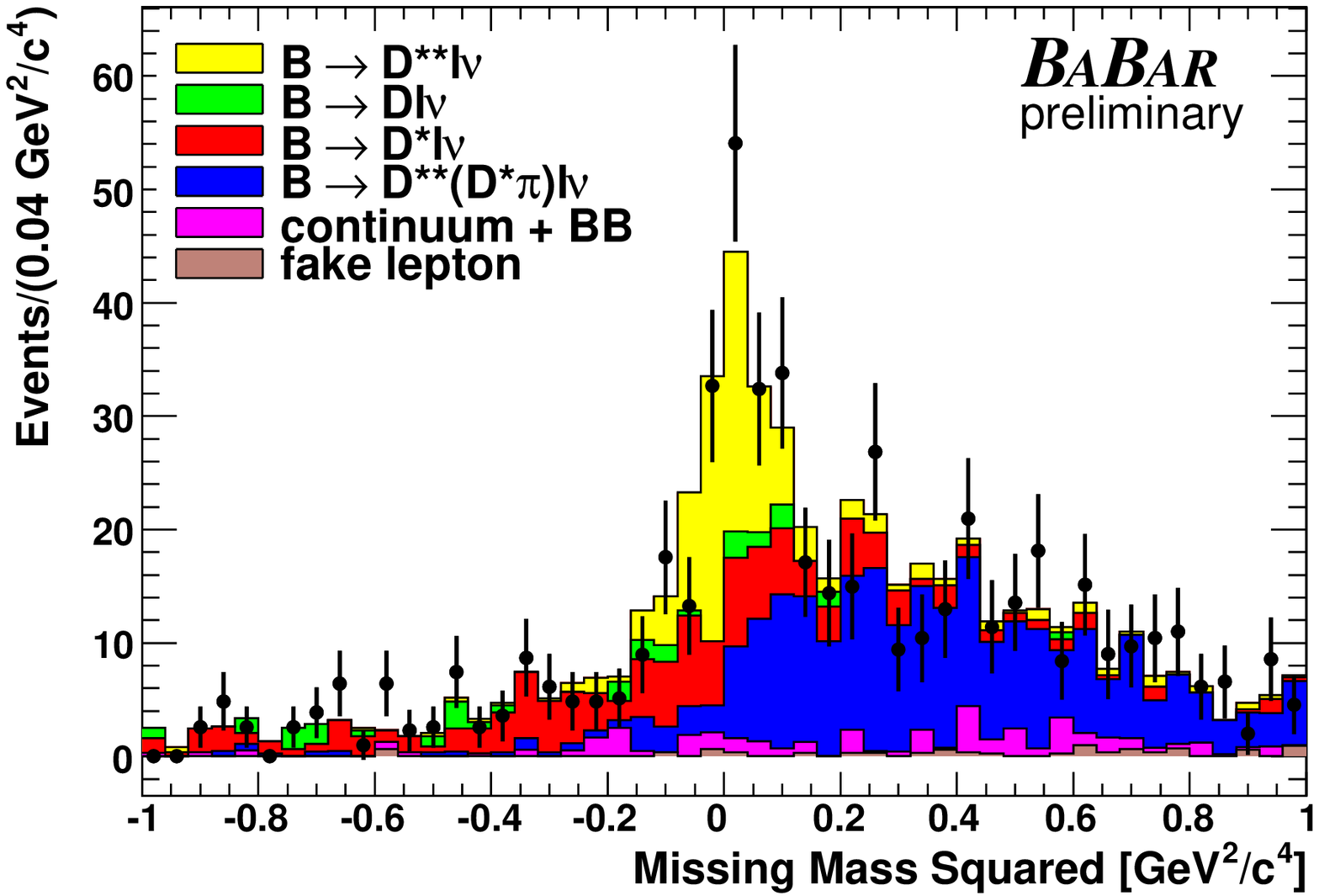,width=8.1cm}
\epsfig{figure=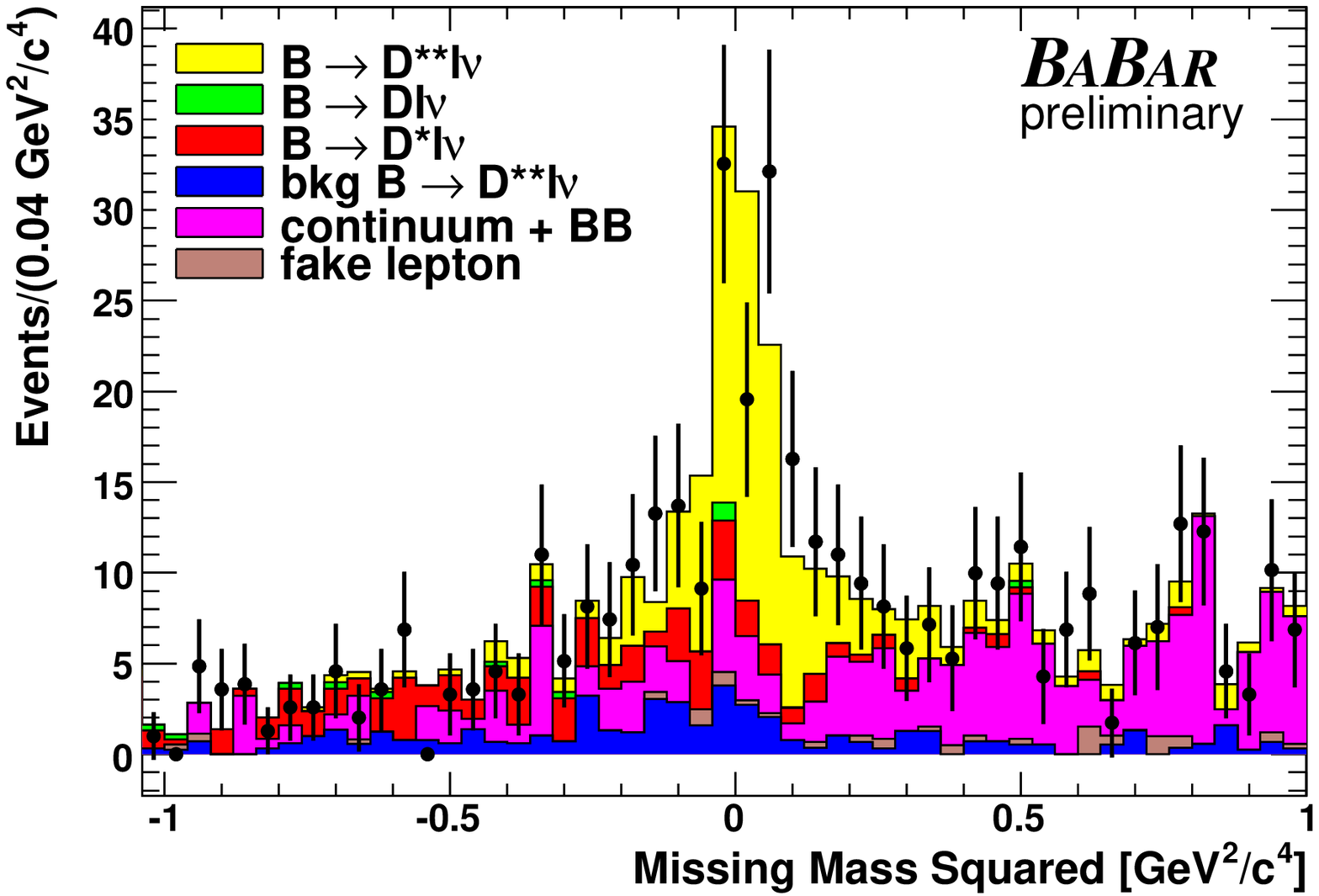,width=8.1cm}
\caption{Fit to the $m^2_{miss}$ distribution for $B^- \to D^+ \pi^- \ell^- \bar{\nu}_{\ell}$ (top left), $B^- \to D^{*+} \pi^- \ell^- \bar{\nu}_{\ell}$ (top right), $\overline{B^0} \to D^0 \pi^+ \ell^- \bar{\nu}_{\ell}$ (bottom left), and $\overline{B^0} \to D^{*0} \pi^+ \ell^- \bar{\nu}_{\ell}$ (bottom right): the data (points with error bars) are compared to the results of the overall fit (sum of the solid histograms). The PDFs for the different fit components are stacked and shown in different colors.}
\label{fig:Fit2}
\end{figure}

The $m^2_{miss}$ distributions are compared with the results of the fits in Figure \ref{fig:Fit1} for the $\overline{B} \to D/D^{*} \ell^- \bar{\nu}_{\ell}$ decays and Figure \ref{fig:Fit2} for the $\overline{B} \to D^{(*)}\pi \ell^- \bar{\nu}_{\ell}$ decays. The fitted signal yields and the signal efficiencies, including the $B_{tag}$ reconstruction, are listed in Table \ref{tab:results}.

\begin{table}[!htb]
\centering
\caption{Signal yields and reconstruction efficiencies for the $\overline{B} \to D^{(*)}(\pi) \ell^- \bar{\nu}_{\ell}$ decays. }
\begin{tabular}{|l|c|c|l|c|c|}
\hline
\hline
Decay Mode & $N_{sig}$ & $\epsilon_{sig} (\times 10^{-4})$ & Decay Mode & $N_{sig}$ & $\epsilon_{sig} (\times 10^{-4})$ \\
\hline
$B^- \rightarrow D^0 \ell^- \bar{\nu}_{\ell}$  & 1635 $\pm$ 61 & 1.71 & $B^- \rightarrow D^+ \pi^- \ell^- \bar{\nu}_{\ell}$ & 174 $\pm$ 25 & 1.02\\
\hline
$B^- \rightarrow D^{*0} \ell^- \bar{\nu}_{\ell}$  & 3050 $\pm$ 73 & 1.27 & $B^- \rightarrow D^{*+} \pi^- \ell^- \bar{\nu}_{\ell}$ & 306 $\pm$ 27 & 1.26\\
\hline
$\overline{B^0} \rightarrow D^+ \ell^- \bar{\nu}_{\ell}$  & \hspace{0.07cm} 852 $\pm$ 40 & 0.94 & $\overline{B^0} \rightarrow D^0 \pi^+ \ell^- \bar{\nu}_{\ell}$ & 107 $\pm$ 20 & 0.60 \\
\hline
$\overline{B^0} \rightarrow D^{*+}\ell^- \bar{\nu}_{\ell}$  & 2045 $\pm$ 55  & 0.91 & $ \overline{B^0} \rightarrow D^{*0} \pi^+ \ell^- \bar{\nu}_{\ell}$ & 130 $\pm$ 20 & 0.66\\
\hline
\hline
\end{tabular}
\label{tab:results}
\end{table}

\begin{figure}[!h]
\centering
\epsfig{figure=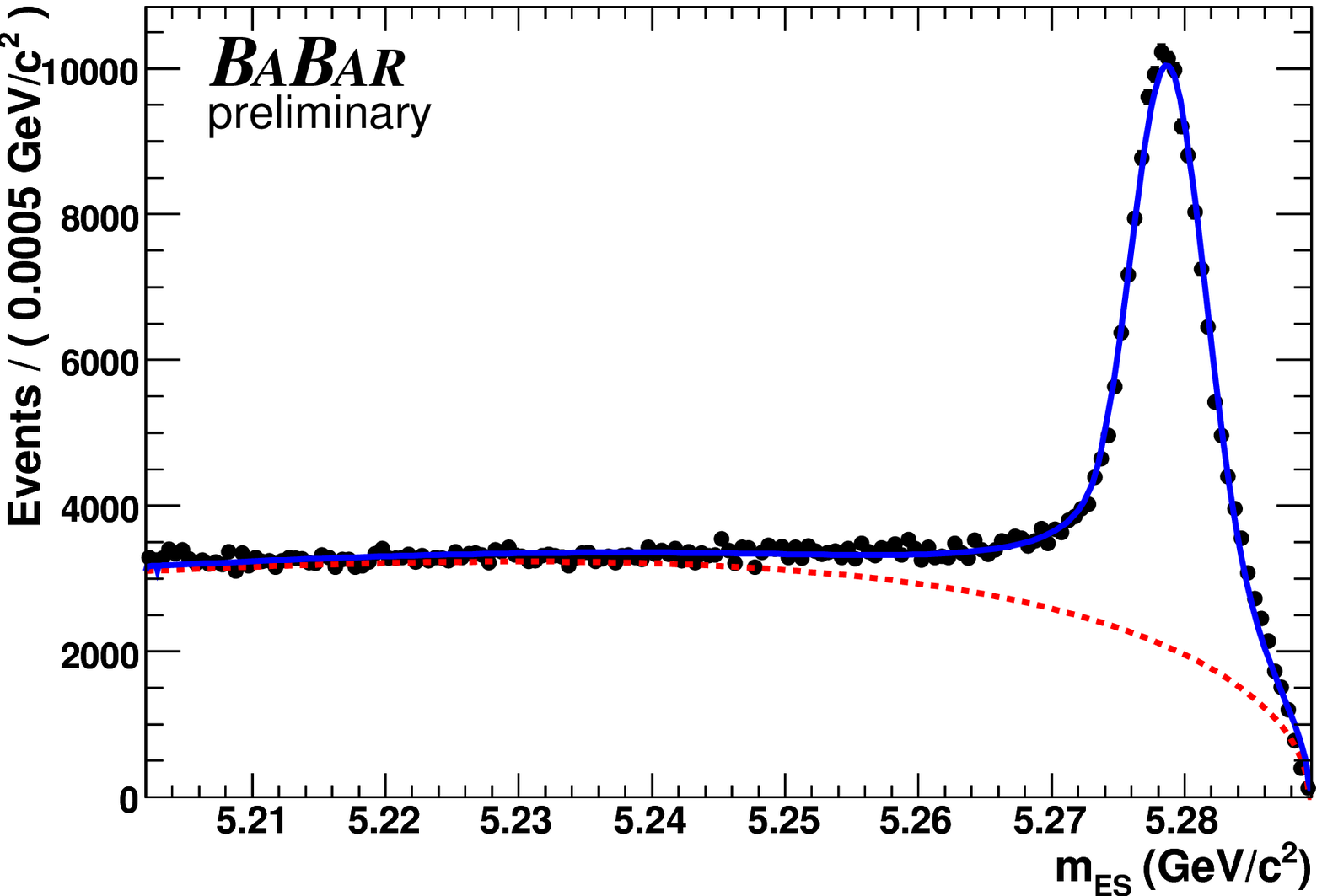,width=8.1cm}
\epsfig{figure=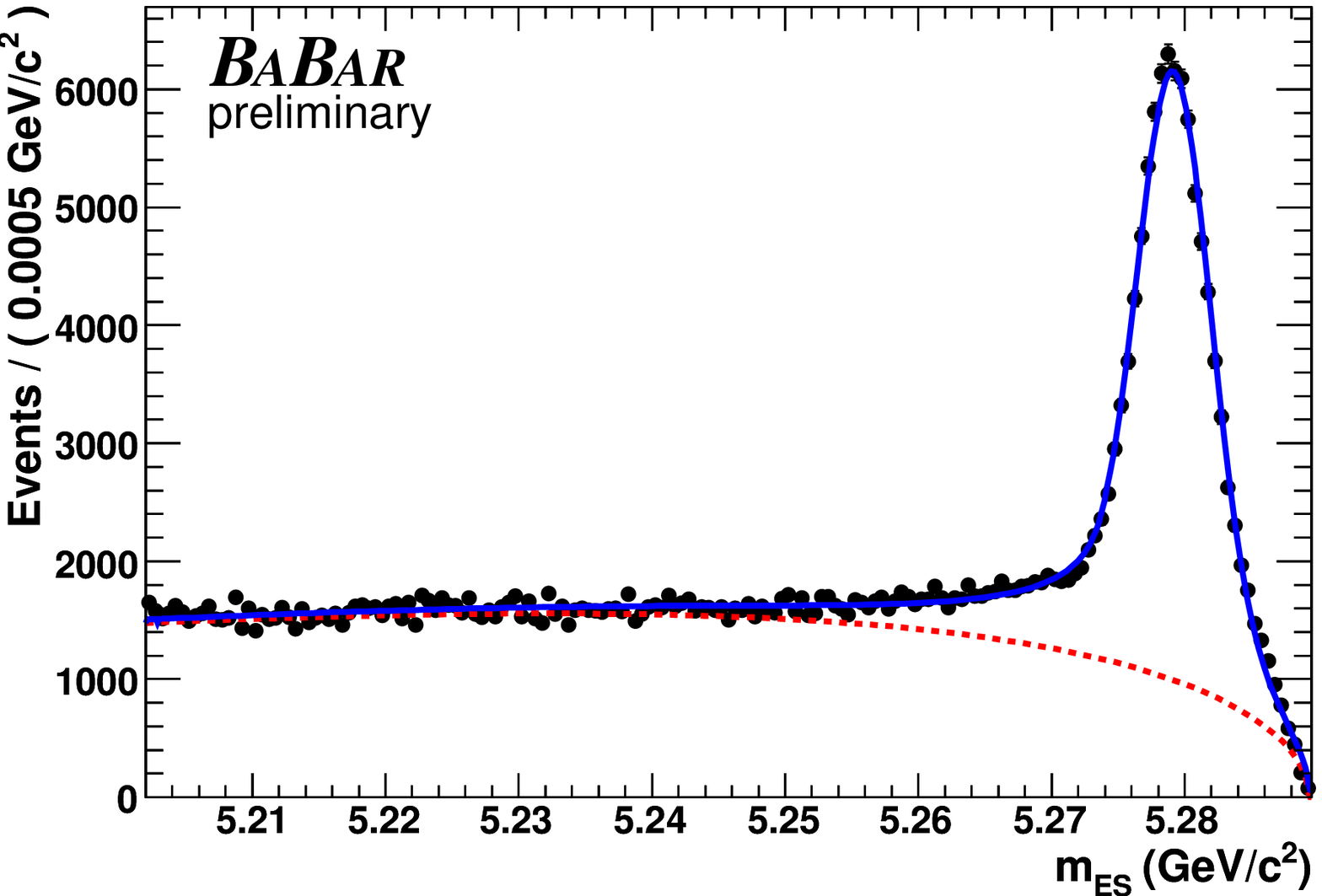,width=8.1cm}
\caption{$m_{ES}$ distributions for the $B^- \to X \ell^- \bar{\nu}_{\ell}$ (left) and $\overline{B^0} \to X \ell^- \bar{\nu}_{\ell}$ (right) sample. The data (points with statistical error) are compared to the result of the fit (solid line). The dashed line shows the contribution of the combinatorial and continuum background.}
\label{fig:mesB}
\end{figure}

In order to reduce the systematic uncertainty, the exclusive ${\cal B} (\overline{B} \rightarrow D^{(*)}(\pi) \ell^- \bar{\nu}_{\ell})$ branching fractions  are measured relative to the inclusive semileptonic branching fraction.
A sample of $\overline{B} \to X \ell^- \bar{\nu}_{\ell}$ events is selected by identifying a charged lepton with CM momentum greater than 0.6 GeV/$c$. In the case of multiple $B_{tag}$ candidates, we select the one reconstructed in the decay channel with the highest purity, defined as the fraction of signal events in the $m_{ES}$ signal region. We require the lepton and the $B_{tag}$ to have the correct charge-flavor correlation and that the lepton track has not been used to reconstruct the $B_{tag}$ candidate. 
Background components peaking in the $m_{ES}$ signal region include cascade $B$-meson decays (i.e. the lepton does not come directly from the $B$) and hadronic decays where one of the hadrons is misidentified as a lepton. These backgrounds are subtracted by using the corresponding simulated Monte Carlo distributions.  The cascade-$B$ meson decays (17.6\% and 19.0\% of the total $m_{ES}$ distribution for charged and neutral $B$, respectively) are reweighted to account for differences between the branching fractions used in our Monte Carlo simulation and the latest experimental measurements~\cite{thorsten}.
The total yield for the inclusive $\overline{B} \to X \ell^- \bar{\nu}_{\ell}$ decays is obtained from a maximum-likelihood fit to the $m_{ES}$ distribution for the $B_{tag}$ candidates using an ARGUS function~\cite{Argus} for the description of the combinatorial \BB\  and continuum background, and a Crystal Ball function~\cite{CrystallBall} for the signal. A broad peaking component is observed in the $m_{ES}$ signal region and is included in the signal definition. This is due to real $\overline{B} \to X \ell^- \bar{\nu}_{\ell}$ decays for which, in the $B_{tag}$ reconstruction, neutral particles have not been identified or have been interchanged with the semileptonic decays (e.g. a $\gamma$ from radiative $D^{*0}$ decay which belongs  to the $D^{*0}$ seed in the $B_{tag}$ decay chain and is instead associated with a $B^- \to D^{*0} (D^0 \gamma) \ell^- \bar{\nu}_{\ell}$ decay).
This broad peaking component is modeled with additional Crystal Ball and ARGUS functions, whose parameters are fixed to the Monte Carlo prediction, except for the Crystal Ball mean value. 
Figure \ref{fig:mesB} shows the $m_{ES}$ distribution for the $B_{tag}$ candidates in the $B^- \to X \ell^- \bar{\nu}_{\ell}$ and $\overline{B^0} \to X \ell^- \bar{\nu}_{\ell}$ sample. The fit gives 159896 $\pm$ 1361 events for the 
$B^- \to X \ell^- \bar{\nu}_{\ell}$ sample and 96771 $\pm$ 968 events for the $\overline{B^0} \to X \ell^- \bar{\nu}_{\ell}$ sample. 

The relative branching fractions ${\cal B} (\overline{B} \rightarrow D^{(*)}(\pi) \ell^- \bar{\nu}_{\ell})/{\cal B} (\overline{B} \to X \ell^- \bar{\nu}_{\ell})$ are obtained by correcting the signal yields obtained from the fit for the reconstruction efficiencies (estimated on \BB\ Monte Carlo events) and using as normalization the inclusive  $\overline{B} \to X \ell^- \bar{\nu}_{\ell}$ signal yield, following the relation 

\begin{displaymath}
\frac{{\cal B}(\overline{B} \rightarrow D^{(*)}(\pi) \ell^- \bar{\nu}_{\ell})}{{\cal B}(\overline{B} \to X \ell^- \bar{\nu}_{\ell})} = \frac{N_{sig}}{\epsilon_{sig}}\frac{\epsilon_{sl}}{N_{sl}}.
\end{displaymath}

\noindent Here, $N_{sig}$ is the number of 
$\overline{B} \to D^{(*)}(\pi) \ell^- \bar{\nu}_{\ell}$ signal events 
for the various modes, reported in  Table~\ref{tab:results} together
 with the corresponding reconstruction efficiencies  
$\epsilon_{sig}$. $N_{sl}$ is the $\overline{B} \to X \ell^- 
\bar{\nu}_{\ell}$ signal yield, and $\epsilon_{sl}$ 
is the corresponding reconstruction efficiency, 
including the $B_{tag}$ reconstruction, 
equal to 0.36\% and 0.23\%, for the $B^- \to X \ell^- \bar{\nu}_{\ell}$ 
and $\overline{B^0} \to X \ell^- \bar{\nu}_{\ell}$ decays, respectively. Using the semileptonic branching fraction ${\cal B}(\overline{B} \to X \ell^- \bar{\nu}_{\ell})= ( 10.78 \pm 0.18)\%$~\cite{pdg} and the ratio of the $B^0$ and the $B^+$ lifetimes $\tau_{B^+}/\tau_{B^0} = 1.071 \pm 0.009$~\cite{pdg}, we compute the absolute branching fractions ${\cal B} (\overline{B} \rightarrow D^{(*)}(\pi) \ell^- \bar{\nu}_{\ell})$.

\section{Systematic Uncertainty}
\label{sec:Systematics}

Different sources of systematic uncertainties have been estimated and are given in Tables \ref{tab:Syst1} and \ref{tab:Syst2}. We have grouped them into several categories.

Detector-related systematics may arise from
differences in the simulation of the track reconstruction and efficiency, particle
identification and neutral particle reconstruction.
The systematic uncertainty related to the reconstruction of charged tracks is determined by randomly removing a fraction of tracks corresponding to the uncertainty in the track finding efficiency, estimated on $e^+e^- \to \tau^+\tau^-$ data control samples.  The systematic uncertainty due to the reconstruction of neutral particles in the EMC is studied by varying the resolution and efficiency to match those found in data control samples.
We estimate the systematic uncertainty due to particle identification by varying the electron and muon identification efficiencies by 2\% and 3\%, respectively. The misidentification probabilities are varied by 15\% for both electrons and muons.
For the modes involving the reconstruction of $D^{*+}$ and $D^{*0}$ states, an additional systematic uncertainty arising from the simulation of low momentum particles is considered and is derived from the measured $m^2_{miss}$ distributions. From the fit to the $\overline{B} \to D X \ell^- \bar{\nu}_{\ell}$ $m^2_{miss}$ distributions in Figure \ref{fig:Fit1}, we extract simultaneously the signal yield of $\overline{B} \to D \ell^- \bar{\nu}_{\ell}$ and $\overline{B} \to D^{*} \ell^- \bar{\nu}_{\ell}$ decays. By comparing the branching fraction for the $\overline{B} \to D^{*} \ell^- \bar{\nu}_{\ell}$ decays obtained from the fit to $m^2_{miss}$ for $\overline{B} \to D \ell^- \bar{\nu}_{\ell}$ and $\overline{B} \to D^{*} \ell^- \bar{\nu}_{\ell}$ events, we evaluate the effect of the soft particle reconstruction. 

We evaluate the systematic uncertainties associated with the Monte Carlo simulation of various signal and background processes. 
We include a systematic uncertainty of 13\%~(19\%) on the photon conversion ($\pi^0$ Dalitz decay) reconstruction efficiency. 
The fraction of $B$ cascade decays in the $\overline{B} \to X \ell^- \bar{\nu}_{\ell}$ sample is varied within its uncertainties and the differences in the $\overline{B} \to X \ell^- \bar{\nu}_{\ell}$ signal yields are included in the systematic uncertainties. 
Possible differences in the $B_{tag}$ composition of the MC simulation and data can affect the efficiencies and the cross-feed between charged and neutral $B$  events. 
To evaluate
this effect, and to account for the limited size of the Monte Carlo sample, we assume a 30\% systematic uncertainty for the cross-feed fractions and we evaluate the systematic
uncertainty by looking at differences in the fitted semileptonic branching
fractions as we change the cross-feed fractions.
We vary the $\overline{B} \to D \ell^- \bar{\nu}_{\ell}$ and  $\overline{B} \to D^{*} \ell^- \bar{\nu}_{\ell}$ form factors within their measured uncertainties~\cite{HQETBaBar} and we include a contribution due to the uncertainties on the branching fractions of the reconstructed $D$ and $D^{*}$ modes, and on the absolute branching fraction ${\cal B} (\overline{B} \to X \ell^- \bar{\nu}_{\ell})$ used for the normalization.

We also evaluate a systematic uncertainty due to differences in the efficiency of the $B_{tag}$ selection in the exclusive selection of $\overline{B} \to D^{(*)}(\pi) \ell^- \bar{\nu}_{\ell}$ decays and the inclusive $\overline{B} \to X \ell^- \bar{\nu}_{\ell}$ reconstruction, by using the same $B_{tag}$ candidate selection adopted in the $\overline{B} \to X \ell^- \bar{\nu}_{\ell}$ reconstruction also for the $\overline{B} \to D^{(*)}(\pi) \ell^- \bar{\nu}_{\ell}$ decays, and taking the difference in the signal yield, corrected by the reconstruction efficiency, as a systematic uncertainty.

The uncertainty in the determination of the $\overline{B} \to X \ell^- \bar{\nu}_{\ell}$ yield is estimated by using an alternative fit method, which is compared to the result of the nominal  $m_{ES}$ fit to estimate the systematic uncertainty. We consider the $m_{ES}$ distribution from the data and the combinatorial \BB\, continuum and other background components (cascade and hadronic $B$ decays) modeled with distributions taken from the Monte Carlo simulation. We fit the background normalization on data in the $m_{ES}$ sideband region, defined by $m_{ES} < 5.265$ GeV/$c^2$. The normalization for the continuum background is fixed to the value obtained from off-peak data. The total background contribution is then subtracted by the total number of events in the $m_{ES}$ distribution to extract the  $\overline{B} \to X \ell^- \bar{\nu}_{\ell}$ signal yield.  The uncertainty in the determination of the $\overline{B} \to D^{(*)}(\pi) \ell^- \bar{\nu}_{\ell}$ yield is estimated by changing the PDFs used to model the different contributions in the $m^2_{miss}$ distribution, e.g. by replacing the continuum PDFs with the corresponding one obtained from off-peak data.

\section{Results}
\label{sec:results}

We measure the following branching fractions:

\begin{eqnarray}
{\cal B} (B^- \rightarrow D^0 \ell^- \bar{\nu}_{\ell})  &=& (2.33 \pm 0.09_{stat.} \pm 0.09_{syst.}) \%  \nonumber \\
{\cal B} (B^- \rightarrow D^{*0} \ell^- \bar{\nu}_{\ell}) &=& (5.83 \pm 0.15_{stat.} \pm 0.30_{syst.}) \% \nonumber \\
{\cal B} (\overline{B^0} \rightarrow D^+ \ell^- \bar{\nu}_{\ell})&=& (2.21 \pm 0.11_{stat.} \pm 0.12_{syst.}) \% \nonumber \\
{\cal B} (\overline{B^0} \rightarrow D^{*+}\ell^- \bar{\nu}_{\ell}) &=& (5.49 \pm 0.16_{stat.} \pm 0.25_{syst.}) \% \nonumber \\
{\cal B} (B^- \rightarrow D^+ \pi^- \ell^- \bar{\nu}_{\ell}) &=& (0.42 \pm 0.06_{stat.} \pm 0.03_{syst.} ) \% \nonumber \\
{\cal B} (B^- \rightarrow D^{*+} \pi^- \ell^- \bar{\nu}_{\ell}) &=& (0.59 \pm 0.05_{stat.} \pm 0.04_{syst.}) \% \nonumber \\
{\cal B} (\overline{B^0} \rightarrow D^0 \pi^+ \ell^- \bar{\nu}_{\ell})&=& (0.43 \pm 0.08_{stat.} \pm 0.03_{syst.}) \% \nonumber \\
{\cal B} (\overline{B^0} \rightarrow D^{*0} \pi^+ \ell^- \bar{\nu}_{\ell}) &=& (0.48 \pm 0.08_{stat.} \pm 0.04_{syst.}) \%. \\ \nonumber
\end{eqnarray}

We compute the total branching fractions of the $\overline{B} \rightarrow D^{(*)} \pi \ell^- \bar{\nu}_{\ell}$ decays assuming isospin symmetry, ${\cal B} (\overline{B} \rightarrow D^{(*)}\pi^0 \ell^- \bar{\nu}_{\ell})=\frac{1}{2} {\cal B} (\overline{B} \rightarrow D^{(*)}\pi^{\pm} \ell^- \bar{\nu}_{\ell})$, to estimate the branching fractions of $D^{(*)} \pi^0$ final states. We obtain:

\begin{eqnarray}
{\cal B} (B^- \rightarrow D^{(*)}\pi \ell^- \bar{\nu}_{\ell}) &=& (1.52 \pm  0.12_{stat.} \pm 0.10_{syst.}) \% \nonumber \\ 
{\cal B} (\overline{B^0} \rightarrow D^{(*)}\pi \ell^- \bar{\nu}_{\ell}) &=& (1.37 \pm  0.17_{stat.} \pm 0.10_{syst.}) \%,  \\ \nonumber 
\end{eqnarray}

\noindent where we assume the systematic uncertainties on the $\overline{B} \rightarrow D \pi \ell^- \bar{\nu}_{\ell}$ and $\overline{B} \rightarrow D^{*} \pi \ell^- \bar{\nu}_{\ell}$ modes to be completely correlated. The accuracy of the branching fraction measurement for the $\overline{B} \rightarrow D^{(*)} \ell^- \bar{\nu}_{\ell}$ decays is comparable to that of the current world average~\cite{pdg}. The result for the $\overline{B} \rightarrow D^{(*)} \pi \ell^- \bar{\nu}_{\ell}$ branching fraction is comparable with the BELLE result~\cite{Abe:2005up}.

In conclusion, preliminary results of the branching fractions of exclusive $B$ semileptonic decays have been obtained for the $\overline{B} \rightarrow D^{(*)} (\pi) \ell^- \bar{\nu}_{\ell}$ modes. 

\section{Acknowledgments}
\label{sec:Acknowledgments}

We are grateful for the 
extraordinary contributions of our \pep2\ colleagues in
achieving the excellent luminosity and machine conditions
that have made this work possible.
The success of this project also relies critically on the 
expertise and dedication of the computing organizations that 
support \babar.
The collaborating institutions wish to thank 
SLAC for its support and the kind hospitality extended to them. 
This work is supported by the
US Department of Energy
and National Science Foundation, the
Natural Sciences and Engineering Research Council (Canada),
Institute of High Energy Physics (China), the
Commissariat \`a l'Energie Atomique and
Institut National de Physique Nucl\'eaire et de Physique des Particules
(France), the
Bundesministerium f\"ur Bildung und Forschung and
Deutsche Forschungsgemeinschaft
(Germany), the
Istituto Nazionale di Fisica Nucleare (Italy),
the Foundation for Fundamental Research on Matter (The Netherlands),
the Research Council of Norway, the
Ministry of Science and Technology of the Russian Federation, and the
Particle Physics and Astronomy Research Council (United Kingdom). 
Individuals have received support from 
the Marie-Curie IEF program (European Union) and
the A. P. Sloan Foundation.

\begin{sidewaystable}
\centering
\caption{Systematic uncertainties (relative errors in \%) in the measurement of $\Gamma(\overline{B} \rightarrow D^{(*)} \ell^- \bar{\nu}_{\ell})/\Gamma(\overline{B} \rightarrow X \ell^- \bar{\nu}_{\ell})$.}
\begin{tabular}{|c|c|c|c|c|}
\hline
\hline
 &\multicolumn{4}{c|}{{\scriptsize Systematic uncertainty on $\Gamma(\overline{B} \rightarrow D^{(*)}\ell^- \bar{\nu}_{\ell}) /\Gamma(\overline{B} \rightarrow X \ell^- \bar{\nu}_{\ell})$} } \\
\hline
& $B^- \rightarrow D^0\ell^-\bar{\nu}_{\ell}$ & $B^- \rightarrow D^{*0}\ell^-\bar{\nu}_{\ell}$ & $\overline{B^0} \rightarrow D^{+}\ell^-\bar{\nu}_{\ell}$ & $\overline{B^0} \rightarrow D^{*+} \ell^-\bar{\nu}_{\ell}$ \\
\hline
Tracking efficiency & 1.4 & 1.2 & 1.4 & 1.5 \\
Neutral reconstruction & 0.7  & 1.9 & 0.5 & 1.1 \\
lepton ID & 0.5  & 0.4  & 0.5 & 0.6 \\
Soft particle efficiency & -  & 1.3 & -  & 1.2\\
\hline
Monte Carlo corrections & \multicolumn{4}{c|}{} \\
\hline
Conversion and Dalitz decay background & 0.04 & 0.07  & 0.06 & 0.05\\
Cascade $\overline{B} \to X \to \ell^-$ decay background  & 0.6 & 0.6 & 1.0 & 1.0\\
$\overline{B^0}-B^-$ cross-feed & 0.2 & 0.3 & 0.2 & 0.3\\
Form factors & 0.4 & 0.8 & 0.4 & 0.8 \\
$D^{(*)}$ branching fractions & 2.3  & 3.5 & 4.1  & 2.7\\
\hline
$\overline{B} \to X \ell^- \bar{\nu}_{\ell}$ branching fraction & 1.9  & 1.9 & 1.9  & 1.9\\
\hline
$B_{tag}$ selection & 0.9 & 1.7 & 1.8 & 1.3\\
\hline
Fit technique & \multicolumn{4}{c|}{}\\
\hline
$\overline{B} \to X \ell^- \bar{\nu}_{\ell}$ yield & 0.5 & 0.5 & 0.9 & 0.9\\
$\overline{B} \rightarrow D^{(*)} \ell^- \bar{\nu}_{\ell}$ yield & 0.6 & 0.4 & 1.2 & 0.4\\
\hline
Total systematic error & 3.7 & 5.2 & 5.4 & 4.5\\
\hline
\hline
\end{tabular}
\label{tab:Syst1}
\end{sidewaystable}

\begin{sidewaystable}
\centering
\caption{Systematic uncertainties (relative errors in \%) in the measurement of $\Gamma(\overline{B} \rightarrow D^{(*)}\pi \ell^- \bar{\nu}_{\ell})/\Gamma(\overline{B} \rightarrow X \ell^- \bar{\nu}_{\ell})$.}
\begin{tabular}{|c|c|c|c|c|}
\hline
\hline
 &\multicolumn{4}{c|}{{\scriptsize Systematic uncertainty on $\Gamma(\overline{B} \rightarrow D^{(*)} \pi \ell^- \bar{\nu}_{\ell}) /\Gamma(\overline{B} \rightarrow X \ell^- \bar{\nu}_{\ell})$} } \\
\hline
&  $B^- \rightarrow D^{+}\pi^- \ell^- \bar{\nu}_{\ell}$ & $B^- \rightarrow D^{*+} \pi^- \ell^- \bar{\nu}_{\ell}$ & $\overline{B^0} \rightarrow D^0 \pi^+ \ell^-\bar{\nu}_{\ell}$ & $\overline{B^0} \rightarrow D^{*0} \pi^+ \ell^- \bar{\nu}_{\ell}$  \\
\hline
Tracking efficiency  & 1.8 & 2.7 & 1.5 & 1.7\\
Neutral reconstruction  & 1.7 & 1.8 & 1.1 & 1.8\\
lepton ID  & 2.3 & 3.0 & 2.6  & 1.8\\
Soft particle efficiency  &  - & 1.2 & -  & 1.3\\
\hline
Monte Carlo corrections & \multicolumn{4}{c|}{} \\
\hline
Conversion and Dalitz decay background  & 0.15 & 0.4 & 0.05 & 0.2\\
Cascade $\overline{B} \to X \to \ell^-$ decay background   & 0.6 & 0.6 & 1.0 & 1.0\\
$\overline{B^0}-B^-$ cross-feed  & 0.2 & 0.3 & 0.2  & 0.3\\
Form factors  & 0.4 & 0.8 & 0.4 & 0.8\\
$D^{(*)}$ branching fractions  & 4.2 & 2.9 & 2.5 & 4.4\\
\hline
$\overline{B} \to X \ell^- \bar{\nu}_{\ell}$ branching fraction & 1.9  & 1.9 & 1.9  & 1.9\\
\hline
$B_{tag}$ selection  & 5.0 & 4.3 & 4.0 & 5.6\\
\hline
Fit technique & \multicolumn{4}{c|}{}\\
\hline
$\overline{B} \to X \ell^- \bar{\nu}_{\ell}$ yield   & 0.5 & 0.5 & 0.9 & 0.9\\
$\overline{B} \rightarrow D^{(*)} \pi \ell^- \bar{\nu}_{\ell}$ yield  & 1.2 & 0.9 & 1.8 & 1.5 \\
\hline
Total systematic error  & 7.7 &  7.3 & 6.4 & 8.4 \\
\hline
\hline
\end{tabular}
\label{tab:Syst2}
\end{sidewaystable}

\end{document}